\documentclass[reprint]{JASA}
\begin{document}
\title[sound field reconstruction]
{Sound Field Reconstruction Using a Compact Acoustics-informed Neural Network}
\author{Fei Ma}                                                            
\author{Sipei Zhao}                            
\email{
Electronic mail: Sipei.Zhao@uts.edu.au}
\affiliation{Center for Audio, Acoustics and Vibration, Faculty of Engineering and IT,
University of Sydney Technology, Ultimo, NSW 2007, Australia.}
\author{Ian S. Burnett}                            
\affiliation{DVC and VP STEM, RMIT University, Melbourne, VIC 3000, Australia}

\date{\today}

\begin{abstract}
Sound field reconstruction (SFR) augments the information of a 
sound field captured by a microphone array. Conventional SFR 
methods using basis function decomposition are straightforward 
and computationally efficient, but may require 
more microphones than needed to measure the sound field. 
Recent studies show that pure data-driven and learning-based methods are promising in
some SFR tasks, but they are usually computationally heavy and may fail to reconstruct
a physically valid sound field.
This paper proposes a compact acoustics-informed neural network 
(AINN) method for SFR, whereby the Helmholtz equation is exploited to regularize the neural network. 
As opposed to pure data-driven approaches that solely rely on  measured sound pressures, the integration of the Helmholtz equation improves robustness of the neural network against variations during  the measurement processes and prompts the 
generation of physically valid reconstructions.
The AINN is designed to be compact, and is able to predict not only the sound pressures but also sound pressure gradients within a spatial region of interest based on measured sound pressures along the boundary. 
Numerical experiments with acoustic transfer functions measured in different environments demonstrate the superiority of the AINN method over the traditional cylinder harmonic decomposition
and the singular value decomposition methods. 
\end{abstract}
\maketitle
\section{Introduction}
Microphone arrays are commonly used 
for measuring a sound field~\cite{benesty2008microphone} and 
to maximize the information about the source, a large-aperture array  
with densely-spaced microphones, is preferred for sound field measurements. 
However, that is not always possible due to practical considerations such 
as cost and microphone arrangement~\cite{rafaely2015fundamentals}. 
This necessitates sound field reconstruction (SFR) \cite{william1999,zhang2008iterative,
fernandez2016sound}, a task that aims to reconstruct a sound field apart 
from the limited (sparse) measurements. 

Existing SFR methods can be broadly classified into two categories:
conventional methods based on basis function decomposition and 
recent learning-based methods.
The conventional methods decompose sound field measurements 
into some basis functions, such as cylinder harmonics (CHs)~\cite{william1999}, 
spherical harmonics~\cite{william1999,wabnitz2011time,chen2015theory, verburg2018reconstruction,tang2022wave}, 
prolate spheroidal wave functions~\cite{zhang2023sound}, 
plane waves~\cite{william1999,fernandez2016sound,antonello2017room,schmid2021spatial}, 
and their corresponding weights. 
The basis functions are solutions of the Helmholtz equation~\cite{skudrzyk2012foundations,william1999}, 
the governing partial differential equation (PDE) of acoustic wave propagation, 
and are continuous spatial functions which can be evaluated at arbitrary positions. 
These two factors make the conventional methods easy to compute and generate a physically 
valid reconstruction of the sound field away from the measurement positions. 
However, the basis functions are designed with respect to some coordinate systems~\cite{william1999} 
without considering the statistical characteristics of sound fields. 
Thus, conventional methods may require more than the absolutely necessary number of measurements 
(spatial sampling points) to determine the basis function weights 
to reconstruct a sound field. 
The reality is that the statistical characteristics of a sound field can be used to allow the 
sampling requirement to be relaxed based on singular value decomposition 
(SVD)~\cite{svd1,svd2}, compressive sensing \cite{wabnitz2011time,
verburg2018reconstruction}, statistical learning~\cite{hahmann2021spatial}, or 
Bayesian inference \cite{schmid2021spatial}. 

In contrast to conventional methods, recent learning-based 
methods do not rely on pre-designed basis functions. Instead, 
they exploit  the learned statistical characteristics of sound 
fields for SFR.
Lluıs {\em et al.}~\cite{lluis2020sound} and Kristoersen {\em et al.} \cite{kristoffersen2021deep}  developed U-net-like neural networks, 
which were trained with simulated or measured room impulse responses. 
The U-net-like neural networks achieved superior SFR performance than some of the 
conventional methods in the low frequency range ($<300$ Hz).
Hahmann {\em et al.}~\cite{hahmann2021spatial} proposed to learn basis functions in
local subdomains. The learned basis functions generalized across different rooms and
frequencies, and showed potential for modeling complex sound fields according to 
their local (spatial) or statistical characteristics. 
By further enforcing self-similarity between adjacent local subdomains~\cite{hahmann2022convolutional},
the method attained better SFR performance when few measurements were available.
Most recently, Fernandez-Grande~{\em et al.} examined the use of generative adversarial 
networks for SFR~\cite{fernandez2023generative}. 
The generative adversarial networks recovered some of the sound field energy at high 
frequencies that would otherwise be lost due to under-sampling~\cite{fernandez2023generative}, 
demonstrating the promise of using statistical learning methods to overcome the sampling 
limitations. 
Although the learning-based methods outperformed the conventional methods in some 
SFR tasks, their computations are time-consuming. 
Furthermore, they are purely data-driven and thus do not 
necessarily reconstruct physically valid sound fields~\cite{fernandez2023generative}.

Recently, physical laws have been integrated into neural networks 
for various acoustic studies, such as the Kirchhoff–Helmholtz-based convolutional neural network (CNN) for nearfield acoustic holography~\cite{olivieri2021physics},
the physics-informed CNN (PI-CNN) for sound field estimation~\cite{shigemi2022physics}, 
and the PINN for room impulse response reconstruction~\cite{karakonstantis2023room,
pezzoli2023implicit}.
These studies~\cite{olivieri2021physics,shigemi2022physics,karakonstantis2023room,
pezzoli2023implicit} attempted to reconstruct the sound field within a region 
of interest (ROI) with few measurements inside the ROI. 

In an alternative approach, this paper proposes an acoustic-informed neural 
network (AINN) to reconstruct the sound field within the ROI based on the 
sound pressures measured on its boundary alone~\cite{raissi2019physics,cuomo2022scientific}. 
The AINN is designed to approximate the sound field at the measurement 
positions and is guided by the Helmholtz equation to generate physically valid 
reconstructions away from the measurement positions. 
The AINN is compact and lightweight, making it easier to train than 
large neural networks. 
In addition, owing to the automatic-differentiation of  deep-learning 
libraries~\cite{pezzoli2023implicit}, the AINN is able to reconstruct the 
pressure gradient within the ROI.
Numerical experiments with transfer functions measured with two microphone arrays 
in three different rooms~\cite{zhao2022room} are conducted to compare the proposed 
method with the CH~\cite{william1999} and the SVD~\cite{svd2} methods.
The experimental results demonstrate the superiority of the proposed AINN 
method over the existing methods. 

The remainder of this paper is organized as follows. The problem is formulated in Sec.~\ref{sec:problem}. 
The CH method and the SVD method are reviewed in Sec.~\ref{sec:methodology},
followed the proposed AINN method. 
Numerical experiments are presented in Sec.~\ref{sec:experiment} to validate 
the performance of the proposed AINN method, in comparison to the CH and the 
SVD methods. 
Sec.~\ref{sec:conclusion} concludes this work. 

\section{Problem formulation \label{sec:problem}}
\begin{figure}[t]
\centerline{\includegraphics[trim={0cm 2cm 0 0},clip,width=7cm]{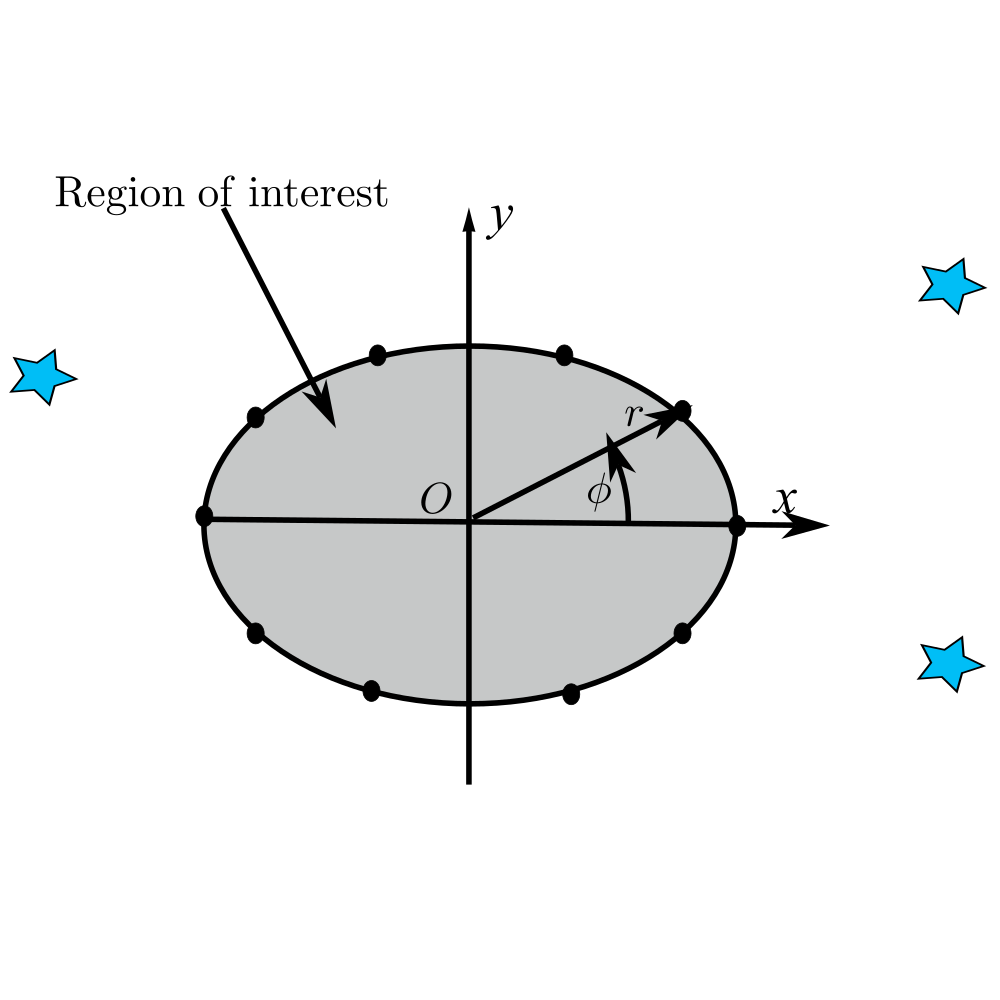}}
\caption{(color online) Problem setup: A number of sources $\star$ generate a sound 
field within a ROI, which is measured by microphones $\bullet$ on the boundary of 
the ROI. The objective is to reconstruct the sound pressure and its gradient within 
the ROI based on the microphone measurement.} 
\label{fig:figure1}
\end{figure}
The problem of interest is illustrated in Fig.~\ref{fig:figure1}, where $(x,y)$ 
and $(r,\phi)$ denote the Cartesian and polar coordinates with respect to the 
origin $O$, respectively. The stars denote sound sources that generate 
the sound field in the Region of Interest (ROI) depicted as the gray area. 
An array of microphones on the boundary of the ROI, 
shown as the dots in Fig. 1, measure the sound pressures at 
$\{x_q,y_q\}_{q=1}^{Q}$ (or $\{r_q,\phi_q\}_{q=1}^{Q}$)
as $P(\omega,x_q,y_q)_{q=1}^{Q}$ (or $P(\omega,r_q,\phi_q)_{q=1}^{Q}$), 
where $\omega=2\pi{f}$ is the angular frequency with $f$ being the frequency. 
The objective is to estimate the sound pressures and their gradients 
inside the ROI based on the measured sound pressures on the boundary. Hereafter, the symbol $\omega$ is omitted in some quantities for notational simplicity.

There is no sound source inside the ROI, and 
hence the sound field within 
the ROI is governed by the homogeneous Helmholtz equation~\cite{william1999}, i.e., 
\begin{equation}
\label{eq:helmholtz_2d}
P+ \frac{1}{(\omega/c)^2} \nabla^2 P =0,
\end{equation}
where $c=340$ m/s is the speed of sound, and 
$\nabla^2$ denotes the Laplacian operator.
In Cartesian coordinates, the Laplacian operator is given by~\cite{william1999}
\begin{equation}
\label{eq:laplacian_car}
\nabla^2 = 
\frac{\partial}{\partial{x^2}}  +   
\frac{\partial}{\partial{y^2}}  , 
\end{equation}
and in polar coordinates it is given by
\begin{equation}
\label{eq:laplacian_polar}
\nabla^2=
\frac{\partial^2 }{\partial r^2} + \frac{1}{r} \frac{\partial }{\partial r} 
+ \frac{1}{r^2} \frac{\partial^2}{\partial \theta^2}. 
\end{equation}

In this paper, we build a compact neural network informed by 
the Helmholtz equation, Eq.~\eqref{eq:helmholtz_2d}, to 
reconstruct the sound field within the ROI based on the 
microphone measurements. 

\section{Methodology\label{sec:methodology}}
This section first reviews the CH method~\cite{william1999} 
and the SVD method~\cite{svd2} for SFR, and subsequently 
proposes the AINN method. 

\subsection{The cylinder harmonics method}
Acoustic quantities are expressed in polar coordinates for the ease of 
computing the CHs. Sound pressures can be decomposed into CHs as~\cite{william1999} 
\begin{equation}
\label{eq:CH_first}
\mathbf{P}_\mathrm{M}=\mathbf{J}\mathbf{A}, 
\end{equation}
where 
$\mathbf{P}_\mathrm{M}=[ 
P(r_1,\phi_1),  
P(r_2,\phi_2),  
...,  
P(r_Q,\phi_Q) ]^{\intercal}$ denote the measured sound pressures at $\{r_q,\phi_q\}_{q=1}^{Q}$,
$(\cdot)^\intercal$ is the transpose operation,
$\mathbf{A}=[ 
A_{-N},
A_{-N+1}, 
..., 
A_{N}
]^{\intercal} $ 
denote the CH weights, 
and 
\begin{equation}
\mathbf{J}=
\begin{bmatrix}
\label{eq:ch_mtx}
J_{-N}(\frac{\omega}{c}r_1)e^{iN\phi_1} &  
... &
J_N(\frac{\omega}{c}r_1)e^{-iN\phi_1}  \\ 
J_{-N}(\frac{\omega}{c}r_2)e^{iN\phi_2}
&
... &
J_N(\frac{\omega}{c}r_2)e^{-iN\phi_2}
\\ 
... & 
... & 
... \\
J_{-N}(\frac{\omega}{c}r_Q)e^{iN\phi_Q}
& 
... &
J_N(\frac{\omega}{c}r_Q)e^{-iN\phi_Q}
\\ 
\end{bmatrix},
\end{equation}
is a $Q\times{(2N+1)}$ matrix whose entry $J_n(\frac{\omega}{c}r_q)e^{-in\phi_q}$ is the $n$-order 
cylinder harmonic~\cite{william1999} evaluated at $(r_q, \phi_q)$, $i$ is the imaginary number unit, 
and $J_n(\cdot)$ is the Bessel function of order $n$~\cite{william1999}. 
In Eqs.~\eqref{eq:CH_first} and \eqref{eq:ch_mtx}, $N$ is the dimensionality of the sound field 
under CH decomposition and is normally chosen as~\cite{kr} 
\begin{equation}
\label{eq:order}
N=\lceil{2\pi{f}r/c}\rceil,
\end{equation}
where $\lceil{\cdot}\rceil$ is the ceiling operation, 

The CH method estimates the weights 
$\hat{\mathbf{A}}=[ 
\hat{A}_{-N},
\hat{A}_{-N+1}, 
..., 
\hat{A}_{N}
]^{\intercal} $ 
through
\begin{equation}
\label{eq:regulation}
\hat{\mathbf{A}}=\mathbf{J}^{\dagger}\mathbf{P}_\mathrm{M}, 
\end{equation}
where $(\cdot)^{\dagger}$ denotes the pseudo-inverse operation. 
The sound pressure and the radial gradient for an arbitrary position $(r_e,\phi_e)$ can be 
reconstructed as~\cite{william1999}
\begin{equation}
\label{eq:ch_sure}
\hat{P}_{\mathrm{CH}}(r_e,\phi_e)
\approx
\sum_{n=-N}^{N}
\hat{A}_{n}
J_{n}(\frac{\omega}{c}r_e)e^{-in\phi_e},
\end{equation}
and 
\begin{equation}
\frac{\partial
\hat{P}_{\mathrm{CH}}(r_e,\phi_e)}{\partial{r_e}}   
\approx\frac{\omega}{c}
\sum_{n=-N}^{N}
\hat{A}_{n}J_{n}^{\prime}(\frac{\omega}{c}r_e)e^{-in\phi_e}, 
\label{eq:ch_rad_gradient}
\end{equation}
where $J_{n}^{\prime}(\cdot)$ denotes the derivative of the Bessel function 
with respect to the argument. 
The pressure gradient along the $x$-axis and the $y$-axis at $(x_e, y_e)$ can be reconstructed as
\begin{equation}
\frac{\partial
\hat{P}_{\mathrm{CH}}(x_e,y_e)}{
\partial{x_e}
} 
=
\frac{ \hat{P}_{\mathrm{CH}}(r_e,\phi_e)}{\partial{r_e}}   
\cos(\phi_e), 
\end{equation}
\begin{equation}
\label{eq:ch_xy_gradient}
\frac{\partial
\hat{P}_{\mathrm{CH}}(x_e,y_e)}{
\partial{y_e}
} 
=
\frac{ \hat{P}_{\mathrm{CH}}(r_e,\phi_e)}{\partial{r_e}}   
\sin(\phi_e).
\end{equation}

\subsection{The SVD method}
The SVD method~\cite{svd1,svd2} regards a source at $(x_s,y_s)$ as a cluster 
of virtual point sources whose positions are $\{x_{s,j},y_{s,j}\}_{j=1}^{J}$ 
and constructs two matrices with respect to the virtual point sources. 
The first one is a matrix of transfer functions between the virtual point 
sources and the microphones 
\begin{equation}
\mathbf{H}_{\mathrm{SM}}=
\begin{bmatrix}
H(x_{1},y_{1},x_{s,1},y_{s,1}) &  
... &
H(x_{1},y_{1},x_{s,J},y_{s,J}) \\ 
... & 
... & 
... \\
H(x_{Q},y_{Q},x_{s,1},y_{s,1}) &  
... &
H(x_{Q},y_{Q},x_{s,J},y_{s,J})
\end{bmatrix},\quad\;\;
\label{eq:svd_first}
\end{equation}
where the $q$-th row and $j$-th column entry, $H(x_{q},y_{q},x_{s,j},y_{s,j})$,
is the free-field transfer function~\cite{william1999} between the virtual point source located at 
$(x_{s,j},y_{s,j})$ and the microphone located at $(x_{q},y_{q})$~\cite{william1999}.
The second one is a matrix of transfer functions between the virtual point sources 
and the pressure estimation points   
\begin{equation}
\mathbf{H}_{\mathrm{SV}}=
\begin{bmatrix}
H(x_{1},y_{1},x_{s,1},y_{s,1}) &  
... &
H(x_{1},y_{1},x_{s,J},y_{s,J}) \\ 
... & 
... & 
... \\
H(x_{V},y_{V},x_{s,1},y_{s,1}) &  
... &
H(x_{V},y_{V},x_{s,J},y_{s,J})
\end{bmatrix},\quad
\end{equation}
where the $v$-th row and $j$-th column entry, $H(x_{v},y_{v},x_{s,j},y_{s,j})$, 
is the free-space transfer function~\cite{william1999} between the virtual point 
source located at $(x_{s,j},y_{s,j})$ and the pressure estimation point located 
at $(x_{v},y_{v})$.
The two matrices are decomposed as~\cite{svd2}
\begin{equation}
\mathbf{H}_{\mathrm{SM}}=\mathbf{U}_{\mathrm{SM}}
\mathbf{\Sigma}_{\mathrm{SM}}\mathbf{V}_\mathrm{SM}^{*},  
\end{equation}
and
\begin{equation}
\mathbf{H}_{\mathrm{SV}}=\mathbf{U}_{\mathrm{SV}}
\mathbf{\Sigma}_{\mathrm{SV}}\mathbf{V}_\mathrm{SV}^{*},    
\end{equation}
where 
$(\cdot)^{*}$ denotes the complex conjugate operation; 
$\mathbf{U}_\mathrm{SM}$ and $\mathbf{U}_\mathrm{SV}$ are unitary matrices whose 
columns represent the basis functions of the receiver space; 
$\mathbf{V}_\mathrm{SM}$  and  $\mathbf{V}_\mathrm{SV}$ are unitary matrices whose 
columns represent the basis functions of the source space; 
and $\mathbf{\Sigma}_\mathrm{SM}$ and $\mathbf{\Sigma}_\mathrm{SV}$ are diagonal 
matrices whose elements represent the capability of a source space basis function to 
excite a receiver space basis function.

The SVD method~\cite{svd2} reconstructs the sound pressure
$\hat{\mathbf{P}}_\mathrm{V}=[\hat{P}(x_1,y_1),...,\\ \hat{P}(x_v,y_v), ...,\hat{P}(x_V,y_V) ]^{\intercal}$ 
as 
\begin{equation}
\hat{\mathbf{P}}_\mathrm{V} 
= 
\mathbf{U}_\mathrm{SV}\mathbf{\Sigma}_{\mathrm{SV}} \mathbf{V}_\mathrm{SV}^{*}
\mathbf{V}_\mathrm{SM}\mathbf{\Sigma}_{\mathrm{SM}}^{-1} \mathbf{U}_\mathrm{SV}^{*}
\mathbf{P}_{\mathrm{M}},
\end{equation}
where $(\cdot)^{-1}$ denotes the matrix inversion operation. 
To reconstruct the pressure gradient, the SVD method first reconstructs the 
pressure at closely spaced positions $(x_v\pm\delta_x,y_v\pm\delta_y)$, and then 
approximates the pressure gradient by 
\begin{equation}
\frac{\partial\hat{P}_{\mathrm{SVD}}(x_v,y_v)}
{\partial{x_v}}
\approx
\frac{
\hat{P}(x_v+\delta_x,y_v)- 
\hat{P}(x_v-\delta_x,y_v) 
}
{2\delta_{x}}, 
\end{equation}
\begin{equation}
\label{eq:svd_xy_gradient}
\frac{\partial\hat{P}_{\mathrm{SVD}}(x_v,y_v)}
{\partial{y_v}}     
\approx
\frac{
\hat{P}(x_v,y_v+\delta_y)- 
\hat{P}(x_v,y_v-\delta_y) 
}
{2\delta_{y}}.
\end{equation}
The radial gradient can be reconstructed as
\begin{equation}
\label{eq:svd_rad_gradient}
\frac{\partial{\hat{P}_{\mathrm{SVD}(r_e,\phi_e)}}}{\partial{r_e}}
= \frac{\partial{\hat{P}_{\mathrm{SVD}}(x_e,y_e)}}{\partial{x_e}}\cos(\phi_e) \nonumber\\
+ \frac{\partial{\hat{P}_{\mathrm{SVD}}(x_e,y_e)}}{\partial{y_e}}\sin(\phi_e). 
\end{equation}

\subsection{The acoustic-informed neural network method}
\begin{figure*}[ht]
\centering
\begin{minipage}[b]{.45\linewidth}
\centering
\centerline{\includegraphics[width=7.0cm]{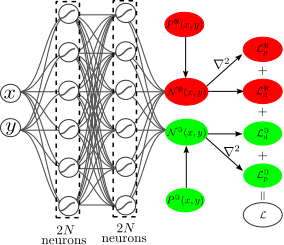}} 
\centerline{(a) cAINN: One coupled network}
\end{minipage}
\quad\quad 
\centering
\begin{minipage}[b]{.45\linewidth}
\centering
\centerline{\includegraphics[width=8.4cm]{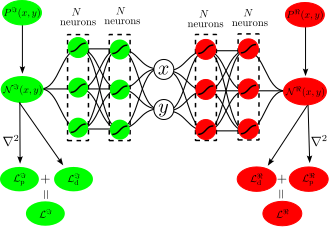}}
\centerline{(b) dAINN: Two decoupled networks}
\end{minipage}
\caption{(color online) Architecture of the AINN: The inputs are Cartesian coordinates, and 
the outputs are the real and imaginary parts of the pressure reconstruction, 
denoted as $\mathcal{N}^{\Re}(x,y)$ and $\mathcal{N}^{\Im}(x,y)$, respectively. 
The data loss and the PDE loss are calculated with respect the pressure 
reconstructions and its Laplacian, respectively.} 
\label{fig:figure2}
\end{figure*}
This section proposes an AINN method for SFR.  
Instead of using complex-valued networks~\cite{complexnn}, whose training can be complicated, to model the frequency-domain 
complex-valued acoustic quantities, we design the AINN method 
using real-valued networks.

Two designs of the AINN, depending on whether the real and imaginary parts 
of the sound pressure are modeled separately 
or collaboratively, are investigated in this paper. 
In the first design, the real and imaginary parts of the sound pressure 
are modeled with a single network and with a single loss function, 
as shown in Fig.~\ref{fig:figure2} (a).  
In the second design, by contrast, the real and imaginary parts of the 
sound pressure are modeled separately with two small networks with 
individual loss functions, as illustrated in Fig.~\ref{fig:figure2} (b). 
Hereinafter, the two designs are referred to as the coupled AINN (cAINN) 
and decoupled AINN (dAINN), respectively. 


To model both the real and imaginary parts of the sound pressure, 
the one network design, cAINN, should have more expressive power and 
hence more hidden layers or more neurons in each hidden layer~\cite{goodfellow2016deep}. 
As shown in  Fig.~\ref{fig:figure2} (a) and (b), this will makes the 
inter-connection between neurons of the one network design
more complicated than that of the two network design, dAINN. 
Although in theory cAINN is capable of exploiting both the real and 
imaginary parts of the sound pressure for the training process and 
hence may achieve better SFR performance, its training is complicated 
and may not achieve the desired performance in practice. 

As shown in Fig.~\ref{fig:figure2},
for both cAINN and dAINN, the inputs are the Cartesian coordinates $(x,y)$ and the outputs are the real and imaginary parts of the reconstructed sound pressure, denoted as $\mathcal{N}^{\Re}(x,y)$ and $\mathcal{N}^{\Im}(x,y)$, respectively. 
Similarly to the conventional data-driven methods, a data loss is utilized to minimize the differences between the reconstructed and the measured sound pressures at the measurement locations. The data loss for the real parts of the sound pressures $\mathfrak{L}_\mathrm{d}^{\Re}$ is given by 
\begin{equation}
\label{eq:data_re_loss}
\mathcal{L}_\mathrm{d}^{\Re}
= \frac{1}{Q}
\sum_{q=1}^{Q}(
P^{\Re}(x_q,y_q)-
\mathcal{N}^{\Re}(x_q,y_q))^2,
\end{equation}
where $P^{\Re}(x_q,y_q)$ and $\mathcal{N}^{\Re}(x_q,y_q)$ denote 
the real parts of the measured and reconstructed sound pressure at 
$\{x_q,y_q\}_{q=1}^{Q}$, respectively.  

To incorporate the acoustic information into the design 
of the neural network, 
the Helmholtz Eq. (1) is utilized to calculate an extra partial 
differential equation (PDE) loss. It is noted that, different from the 
data loss that is calculated only for the measured  locations, the PDE 
loss is calculated for both the measured locations on the boundary and 
the SFR locations within the ROI. By uniformly sampling ${D}$ 
positions within the ROI at $\{x_d,y_d\}_{d=1}^{D}$ and referring 
to Eqs.~\eqref{eq:helmholtz_2d} and \eqref{eq:laplacian_car}, 
the PDE loss for the real part of the sound pressure 
$\mathfrak{L}_\mathrm{p}^{\Re}$ is given by
\begin{equation}
\label{eq:pde_re_loss}
\begin{split}
\mathcal{L}_\mathrm{p}^{\Re}
=&
\frac{1}{D}
\sum_{d=1}^{D}
\Big( 
\mathcal{N}^{\Re}(x_d,y_d)
\nonumber\\
&+
\frac{1}{(w/c)^2}\Big [\frac{\partial^2{}
\mathcal{N}^{\Re}(x_d,y_d)
} {\partial{x_d^2}}
+
 \frac{\partial^2{}
\mathcal{N}^{\Re}(x_d,y_d)
 } {\partial{y_d^2}} \Big ] 
\Big)^2,             \quad                                
\end{split}
\end{equation}
where $\mathcal{N}^{\Re}(x_d,y_d)$ denotes the real-part of the  pressure reconstruction  at $\{x_d,y_d\}$. The data loss and the PDE loss are combined and results in the total loss. 
The definitions of the imaginary-part data loss $\mathcal{L}_\mathrm{d}^{\Im}$ 
and PDE loss $\mathcal{L}_\mathrm{p}^{\Im}$ are similar to Eqs.~\eqref{eq:data_re_loss}
and \eqref{eq:pde_re_loss}, respectively, and are not shown for brevity. 

For the cAINN in Fig.~\ref{fig:figure2} (a), a single network is used to model 
the real and imaginary parts of the sound pressure, and there are $2N$ 
neurons in each hidden layer. 
The trainable parameters of the network are updated to minimize a single total loss 
function 
\begin{equation}
\label{eq:tot_loss}
\mathcal{L}
= \mathcal{L}_\mathrm{d}^{\Re} + \mathcal{L}_\mathrm{p}^{\Re} 
+ \mathcal{L}_\mathrm{d}^{\Im} + \mathcal{L}_\mathrm{p}^{\Im}.   
\end{equation}

For the dAINN in Fig.~\ref{fig:figure2} (b), 
two independent networks are used to model the real and imaginary parts 
of the sound pressure, respectively, with $N$ neurons in each hidden 
layer. The trainable parameters of two networks are updated to minimize 
the real-part loss  
\begin{equation}
\label{eq:tot_re_loss}
\mathcal{L}^{\Re}
= \mathcal{L}_\mathrm{d}^{\Re} + \mathcal{L}_\mathrm{p}^{\Re}, 
\end{equation}
and the imaginary-part loss 
\begin{equation}
\label{eq:tot_im_loss}
\mathcal{L}^{\Im}
=  \mathcal{L}_\mathrm{d}^{\Im} + \mathcal{L}_\mathrm{p}^{\Im},   
\end{equation}
respectively. 
Once trained, the AINN method can reconstruct the sound pressure at an arbitrary 
position $(x_e,y_e)$ as $\mathcal{N}(x_e,y_e)=\mathcal{N}^{\Re}(x_e,y_e)+i\mathcal{N}^{\Im}(x_e,y_e)$, 
or equivalent $\mathcal{N}(r_e,\phi_e)$.
The sound pressure gradient at that position can be reconstructed  
as ${\partial{\mathcal{N}(x_e,y_e)}}/{\partial{x_e}}$ along the $x$ direction  
and as ${\partial{\mathcal{N}(x_e,y_e)}}/{\partial{y_e}}$ along the $y$ direction
through differentiation on the network output. 
The pressure gradient along the radial direction can be reconstructed as 
\begin{equation}
\label{eq:ainn_rad_gradient}
\frac{\partial{\mathcal{N}(r_e,\phi_e)}}{\partial{r_e}}
= \frac{\partial{\mathcal{N}(x_e,y_e)}}{\partial{x_e}}\cos(\phi_e) 
+ \frac{\partial{\mathcal{N}(x_e,y_e)}}{\partial{y_e}}\sin(\phi_e). 
\end{equation}

Here are comments on the AINN method and recommended configurations: 
\begin{enumerate}
\item \noindent\textbf{Using $\tanh$ as the activation function:\\}
We use $\tanh$ as the activation function for two reasons. 
First, the $\tanh$ function is a smooth function, whose second-order gradient can 
be computed. This is necessary for the Laplacian operator, Eqs.~\eqref{eq:laplacian_car} and \eqref{eq:laplacian_polar}.  
Second, the $\tanh$ function outputs positive or negative values according to the input.
This makes it easier to model sound pressure, whose value can be either positive or negative. 

\item 
\noindent \textbf{Cartesian coordinates vs polar coordinates:\\}
For the AINN method, we express sound pressures in Cartesian coordinates instead of polar coordinates for two reasons.
First, the presence of the $1/r$ term can make the Laplacian 
operator in polar coordinates, Eq.~\eqref{eq:laplacian_polar}, 
to be numerically unstable. 
Second, for the sound field on a circle,
there is no pressure variation along the radial direction. In the case, the 
AINN method is unable to estimate the first- and second-order radial gradient 
needed for calculating the Laplacian operator in polar coordinates, Eq.~\eqref{eq:laplacian_polar}.

    \item 
\noindent\textbf{Loss function:\\}
The loss functions, Eqs.~\eqref{eq:tot_loss}, \eqref{eq:tot_re_loss}, and 
\eqref{eq:tot_im_loss}, consist of both the data loss and the PDE loss.   
The data loss prompts the network output to approximate the 
measured sound pressure at positions $\{x_q, y_q\}_{q=1}^{Q}$, 
which are on the boundary of the ROI as shown in Fig.~\ref{fig:figure1}. The PDE loss, on the other hand, regularizes the network output to conform with the 
Helmholtz equation at positions $\{x_d, y_d\}_{d=1}^{D}$ both on the boundary 
of and within the ROI.
$D$ should be chosen to be sufficiently large and thus the distance between 
adjacent positions is at most half (ideally one tenth) of the wave-length for 
the frequency of interest.

\item \noindent \textbf{Neuron number:\\}
As shown in Eq.~\eqref{eq:ch_sure}, the sound pressure can be expressed as a linear 
combination of $2N+1$ CHs $\{J_n(kr)e^{-in\phi}\}_{n=-N}^{N}$, 
which are solutions of the Helmholtz equation, Eq.~\eqref{eq:helmholtz_2d}. 
As shown in Fig.~\ref{fig:figure2} (a) and (b), the very same sound pressure can also be expressed as a linear combination of the output of a number of neurons.
This fact inspires us to set to the number of neurons in hidden layers according to the CH decomposition of the sound pressure. 
Specifically, for the cAINN, the one network design, the neuron number 
is set to be $2N$, and for the dAINN, 
the two network design, the neuron number is set to be $N$, 
where  $N=\lceil{2\pi{}f{}r/c}\rceil\approx{}\lceil{fr/50}\rceil$. 
For a sound field whose size approximates the size of human heads, i.e.,
$r\approx0.1$ m, $N\leq20$ for $f\leq10$ kHz.

As shown in Sec.~\ref{sec:experiment}, an AINN with two hidden 
layers and less than 10 neurons on each hidden layers is found 
to be sufficient for modeling the sound fields around two arrays. 
Therefore, the AINN is compact and lightweight in comparison to 
other learning-based methods \cite{lluis2020sound,kristoffersen2021deep,hahmann2021spatial, hahmann2022convolutional,fernandez2023generative, 
olivieri2021physics,shigemi2022physics,karakonstantis2023room,
pezzoli2023implicit}.
\end{enumerate}

\begin{figure*}[ht]
\centering
\begin{minipage}{0.45\textwidth}
\centering
\includegraphics[width=0.9\linewidth]{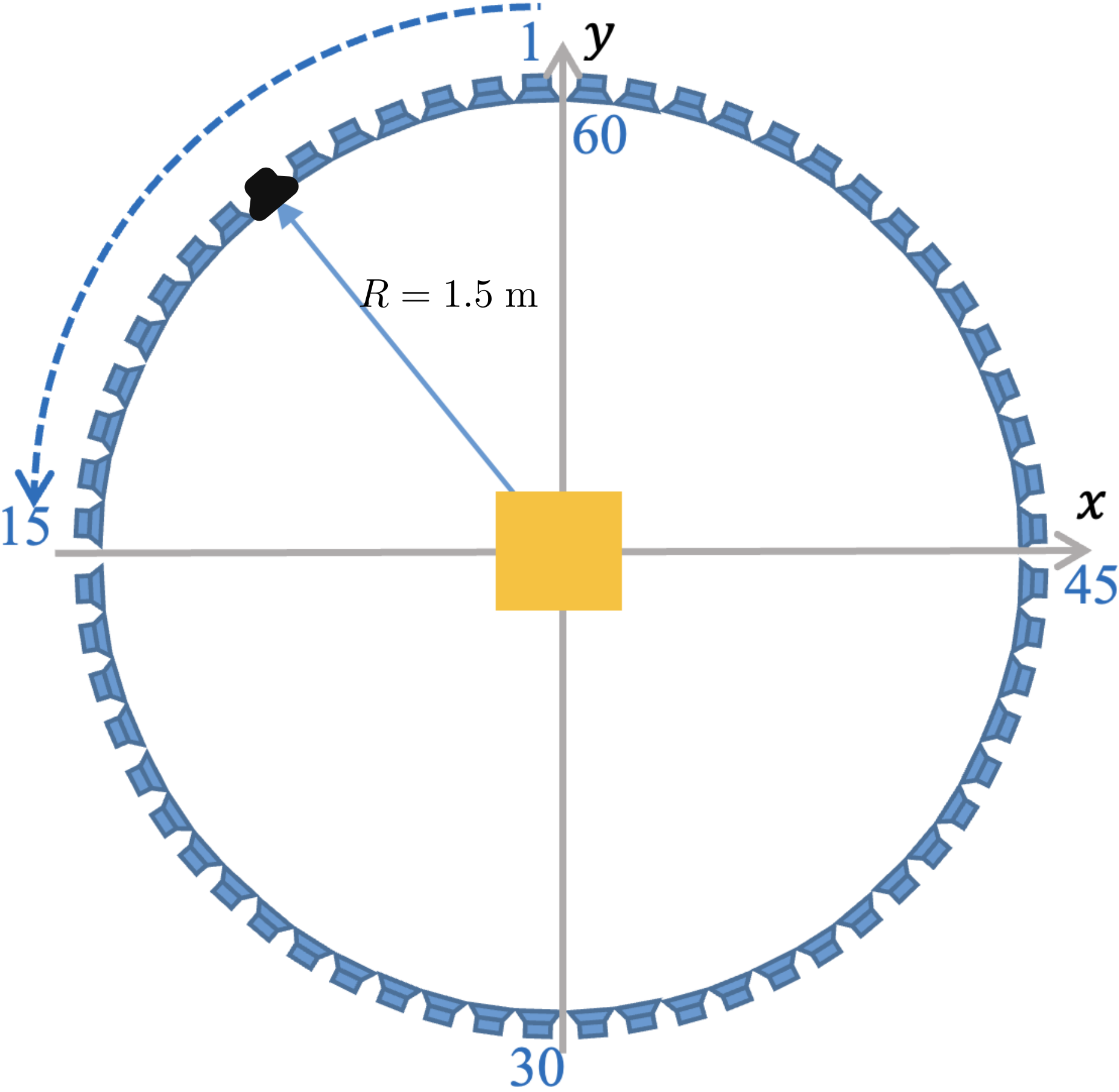}
\centerline{(a) A 60-loudspeaker array with}
\centerline{a microphone array at the origin.}
\end{minipage}%
\qquad 
\begin{minipage}{0.45\textwidth}
\centering
\includegraphics[width=0.85\linewidth]{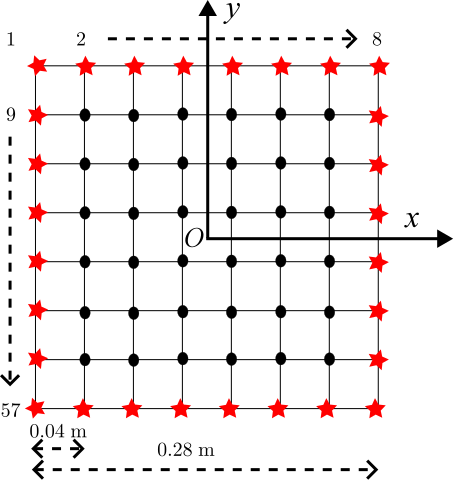}
\centerline{(b) A 64-microphone planar array: 28 }
\centerline{exterior microphones ($\star$) and 36}
\centerline{interior microphones ($\bullet$).}
\end{minipage}%
\qquad 
\begin{minipage}{0.45\textwidth}
\includegraphics[width=0.95\linewidth]{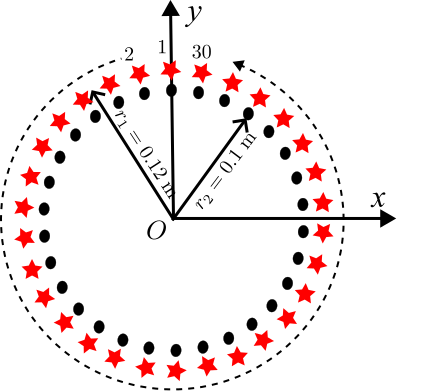}
\centerline{(c) A dual-circular array:}
\centerline{ 30 exterior microphones ($\star$)}
\centerline{ and 30 interior microphones ($\bullet$).}
\end{minipage}
\caption{(color online) Experiment setup~\cite{zhao2022room}: 
the loudspeaker array and the microphone arrays.}
\label{fig:figure3}
\end{figure*}
\begin{figure*}[ht]
\centering
\begin{minipage}[b]{0.31\linewidth}
\centering
\centerline{\includegraphics[width=5.7cm]{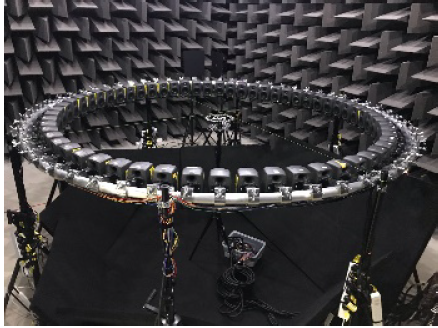}}
\centerline{(a) An anechoic chamber}
\end{minipage}
\quad 
\centering
\begin{minipage}[b]{0.31\linewidth}
\centering
\centerline{\includegraphics[width=5.7cm]{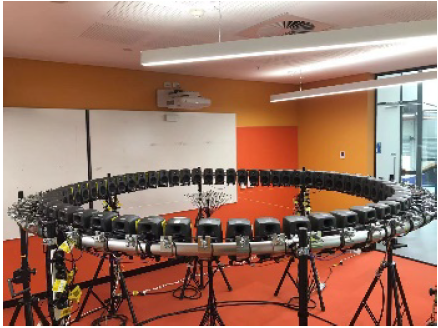}}
\centerline{(b) A medium room}
\end{minipage}
\quad 
\centering
\begin{minipage}[b]{0.31\linewidth}
\centering
\centerline{\includegraphics[width=5.7cm]{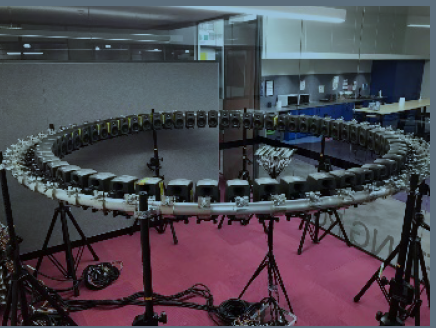}}
\centerline{(c) A small room  }
\end{minipage}
\caption{(color online) Experiment setup: three room environments.}
\label{fig:figure4}
\end{figure*}

\section{Experiments\label{sec:experiment}}
Numerical experiments were conducted to validate the performance 
of the proposed AINN method, and to compare it with the CH 
and SVD methods. 
\subsection{Data processing}
The SFR methods were evaluated using the UTS multi-zone sound field reproduction dataset~\cite{zhao2022room}. 
The measurement set up is shown in Figs.~\ref{fig:figure3} and \ref{fig:figure4}. 
The room impulse responses (RIRs) between a loudspeaker array (Fig.~\ref{fig:figure3} (a)) 
and two microphone arrays (Fig.~\ref{fig:figure3} (b) and (c)) were measured 
in an anechoic chamber (Fig.~\ref{fig:figure4} (a)), a medium room 
(Fig.~\ref{fig:figure4} (b)), and a small room (Fig.~\ref{fig:figure4} (c)). 
The loudspeaker array and the microphone arrays were arranged concentrically.

The loudspeaker array consisted of 60 Genelec 8010A Studio Monitors~\cite{zhao2022room} 
as shown in Fig.~\ref{fig:figure3} (a), and was placed approximately at 
the center of each room as shown in Fig.~\ref{fig:figure4}. 
The loudspeaker positions are
\begin{equation}
x_l = -R \sin((2l-1)\pi/60)\; \mathrm{m},   
\end{equation}
\begin{equation}
y_l = R \cos((2l-1)\pi/60)\; \mathrm{m},  
\end{equation}
where $l=1,2, ..., 60$.

Two microphone arrays were constructed using the DPA 4060 Series Miniature Omni-directional microphones~\cite{zhao2022room}, which 
were calibrated at 1 kHz. 
The first array is a 64 microphone planar array with side 
length of 0.28 m as shown in Fig.~\ref{fig:figure3} (b).
The microphone positions are 
\begin{equation}
x_m = -0.14 + 0.04 \times \mathrm{mod}(m-1,8) \;\mathrm{m},
\end{equation}
\begin{equation}
y_m = 0.14 - 0.04 \times \lfloor{(m-1)}/{8}\rfloor\; \mathrm{m},
\end{equation}
where $\lfloor\cdot\rfloor$ is the floor operation, and $m=1,2, ..., 64$.
The second array is a dual-circular array with 30 microphones uniformly 
placed on each circle as shown in Fig.~\ref{fig:figure3} (c). 
The microphone positions are
\begin{equation}
\theta_m = 2\pi(m-1)/30,  
\end{equation}
\begin{equation}
x_m = -r \times \sin(\theta_m)\; \mathrm{m}, 
\end{equation}
\begin{equation}
y_m = r \times \cos(\theta_m)\; \mathrm{m},
\end{equation}
where $m=1,2, ..., 30$, and
$r=r_1 = 0.12 $ m for the exterior circle and $r=r_2=0.1$ m for the interior circle.

The RIRs were transformed into frequency-domain transfer functions through the discrete Fourier transform~\cite{oppenheim1997signals}, resulting in the sound pressure used for the experiments. 
For more details about the measurement, please refer to~\cite{zhao2022room}.  

\subsection{Implementation}
\textbf{CH method:}  The CH method was implemented based on Eqs.~\eqref{eq:CH_first} - \eqref{eq:ch_xy_gradient}.
Based on Eq.~\eqref{eq:order}, the dimensionalities of the sound 
field within the planar array are $N=3, 6, 8$ for $f=1, 2, 
3$ kHz, respectively; the dimensionalities of the sound field within 
the dual-circular array are $N=3, 5, 7$ for $f=1, 2, 3$ kHz, 
respectively.  

\textbf{SVD method:}  The SVD method was implemented based on Eqs.~\eqref{eq:svd_first} - \eqref{eq:svd_rad_gradient}.
For a source (loudspeaker) located at $(x_s,y_s)$, the virtual point 
sources are uniformly arranged around the source in a 0.1 m $\times$ 
0.1 m square, and the distance between two virtual point sources is 
0.01 m. This amounts to 120 virtual point sources in total.

\textbf{AINN method:}  We used the TensorFlow library, 
and initialized the trainable parameters according to 
the Xavier initialization~\cite{glorot2010understanding}. 
The ADAM algorithm with a learning rate of 0.001 was used 
as the optimizer.
The AINN method was trained for $10^5$ epochs. The neuron number was set based on the dimensionality of the sound field 
under CH decomposition. 
The hidden layer number was set as $1$ for $f=1$ kHz and as $2$ for 
$f=2, 3$ kHz based on a trial-and-error process. 

The measured sound pressures at the RIO boundary and their corresponding 
AINN reconstructions were used for calculating the data loss. 
On the boundary of and within the ROI, we uniformly selected positions 
with a $0.01$ m adjacent distance,  reconstructed the sound 
pressures at these positions, and calculated the PDE 
loss with respect to them. 

\subsection{Performance metrics}
The performance of all methods was evaluated by the reconstruction error 
\begin{equation}
\label{eq:error}
\xi=10\log10\frac{\sum_{e=1}^{E}||P(x_e,y_e)-\hat{P}(x_e,y_e)||_2^2}{\sum_{e=1}^{E}||P(x_e,y_e)||_2^2},    
\end{equation}
where $||\cdot||_2$ denotes the 2-norm,
$P(x_e,y_e)$ and $\hat{P}(x_e,y_e)$ are the ground-truth sound pressure and its reconstruction, 
respectively, and $\{x_e,y_e\}_{e=1}^{E}$ are coordinates of the sound pressure 
reconstruction positions. The pressure gradient reconstruction error was defined 
similarly to Eq.~\eqref{eq:error}.

\subsection{Sound pressure reconstruction: loudspeaker 7\label{sec:pressure_7}}
\begin{figure}[ht]
\centerline{\includegraphics[width=9cm]{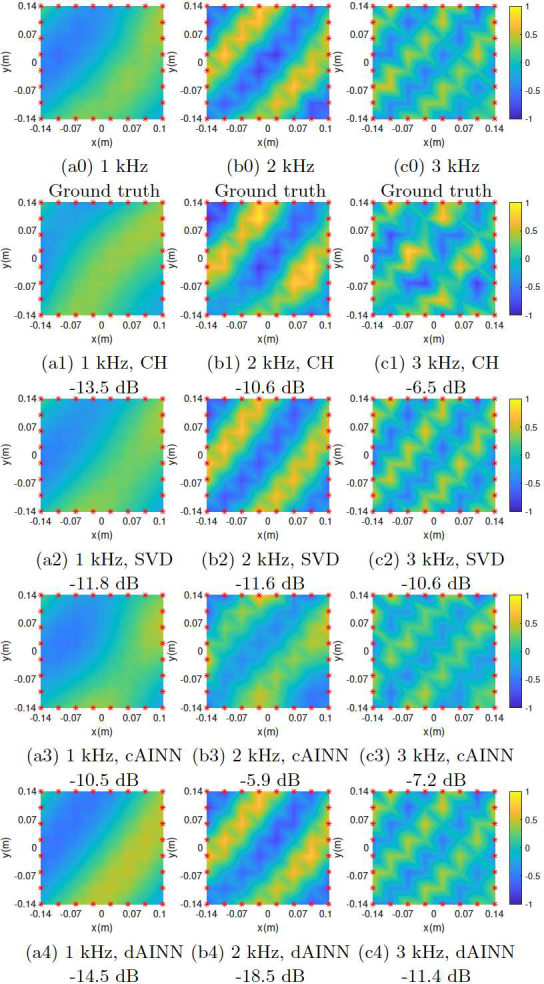}}
\caption{(color online) Pressure reconstruction: reconstruct the pressure within the 
planar array based on the measured pressure ({\color{red}$\star$}) 
- Anechoic chamber.} 
\label{fig:figure5}
\end{figure}

\begin{figure}[ht]
\centerline{\includegraphics[width=9cm]{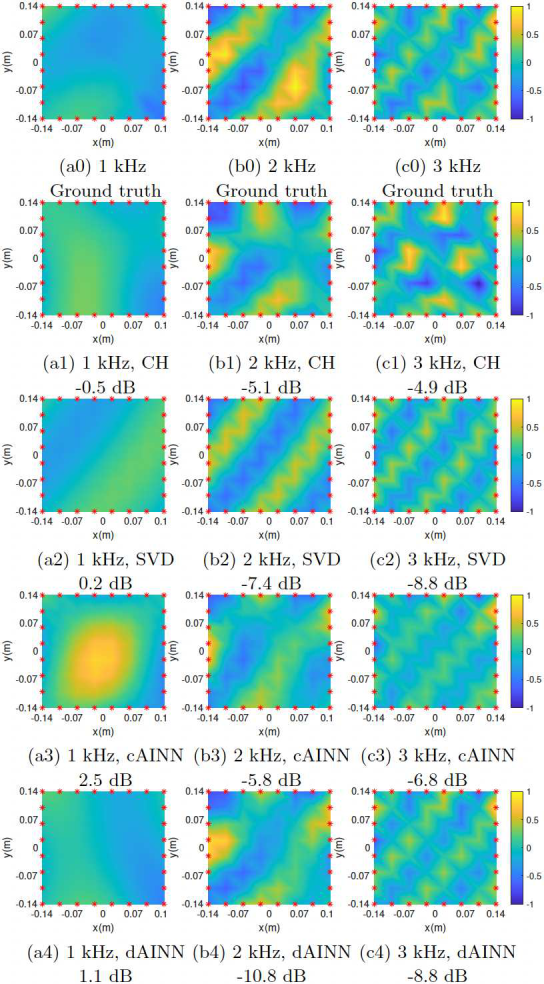}}
\caption{(color online) Pressure reconstruction:
reconstruct the pressure within the planar array based 
on the measured pressure ({\color{red}$\star$}) - medium room.} 
\label{fig:figure6}
\end{figure}

\begin{figure}[ht]
\centerline{\includegraphics[width=9cm]{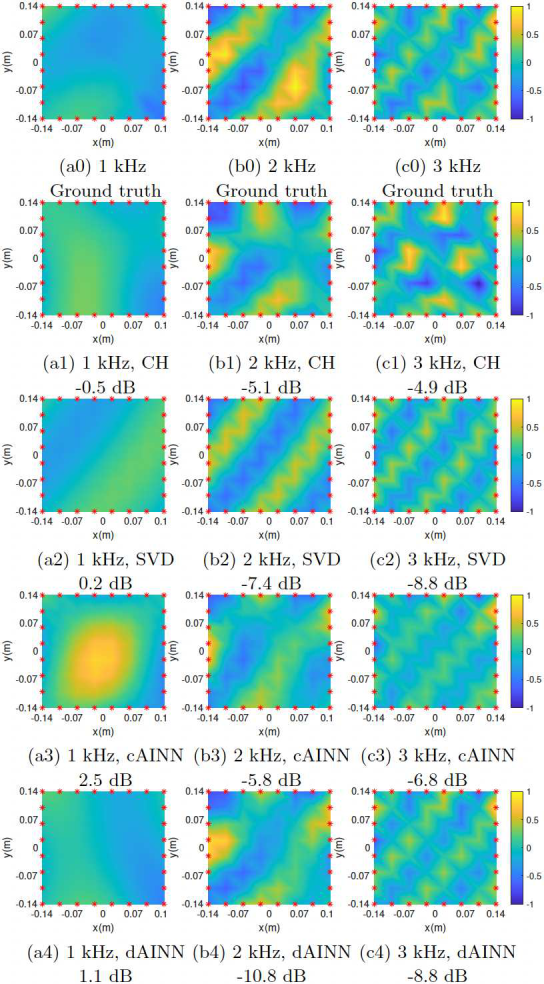}}
\caption{(color online) Pressure reconstruction:
reconstruct the pressure within the planar array based 
on the measured pressure ({\color{red}$\star$}) - small room.} 
\label{fig:figure7}
\end{figure}

Based on the sound pressure measured by the 28 exterior 
microphones of the planar array (Fig.~\ref{fig:figure3} (b)), 
we reconstructed the sound pressure at the 36 interior 
microphones within the array. 
Figs.~\ref{fig:figure5}, \ref{fig:figure6}, and \ref{fig:figure7} 
show the real part of the sound pressure due to loudspeaker 7 (the 
black loudspeaker in Fig.~\ref{fig:figure3} (a))
at 1, 2, and 3 kHz, respectively.
The figures also show the reconstructions by the CH method, 
the SVD method, the AINN method, and corresponding (both real 
and imaginary parts) reconstruction errors. 
The results for the imaginary parts are similar and are not 
shown for brevity. 

\clearpage

\textbf{cAINN vs dAINN:} As shown in Figs.~\ref{fig:figure5}, \ref{fig:figure6}, and \ref{fig:figure7}, 
the dAINN, two network design, outperforms the cAINN, one network design, 
at all frequencies and in all room environments. 
Experiments for the cAINN with more or less neurons on hidden layers, 
i.e., $\lceil{3N/2}\rceil$ or $3N$, were also conducted, and 
the results were also inferior to the dAINN, thus are 
not shown for brevity.  
The results demonstrate that two independent small neural networks 
are better than a single large neural network for modeling 
the real and imaginary parts of the sound pressure. 
Hereinafter, we focus only on the dAINN.

\textbf{dAINN vs CH and SVD at 2 and 3 kHz:} As shown in the second and third columns of Figs.~\ref{fig:figure5}, \ref{fig:figure6}, 
and \ref{fig:figure7}, the SVD method performs better than the CH 
method and in most of the cases for $f=2, 3$ kHz, with a decrease in 
the overall reconstruction error by 1.0 to 2.6 dB at 2 kHz and 1.2 to 4.1 dB 3 kHz across the three rooms. 
This is because the prior information of the sound source location was used 
in the SVD method~\cite{svd1,svd2}. 
In contrast, the dAINN outperforms the SVD method in all the tested rooms 
at $f=2, 3$ kHz, though no prior information about the sound source 
location was required by the dAINN method. 
This could be attributed to the fact that the CH and SVD methods 
relied on only the sound pressures measured on the edge of the planar array for SFR. 
The measured pressures did not necessarily contain sufficient information to 
fully determine the sound field within the planar array. 
The dAINN, on the other hand, exploited the Helmholtz equation for 
regularizing the SFR within the array through the PDE loss and reconstructed 
the sound field within the planar array more accurately. 

\textbf{Pressure variation at 1 kHz:} As shown in the first column of Figs.~\ref{fig:figure5}, \ref{fig:figure6},  and \ref{fig:figure7}, 
at 1 kHz all methods achieve lower than -10 dB 
reconstruction errors in the anechoic chamber,
but not in the medium room and the small room. 
This could be attributed the fact that at 1 kHz in the medium 
(Fig.~\ref{fig:figure6} (a0)) and the small rooms 
(Fig.~\ref{fig:figure7} (a0)), the pressure variations on the edge 
of the planar array were so small that the microphones were unable to capture enough information of the sound field, 
and thus the methods were unable to accurately reconstruct the sound field.

\subsection{Sound pressure reconstruction: all loudspeakers\label{sec:pressure_all}}
\begin{figure}[t]
\centering
\centerline{\includegraphics[width=9.6cm]{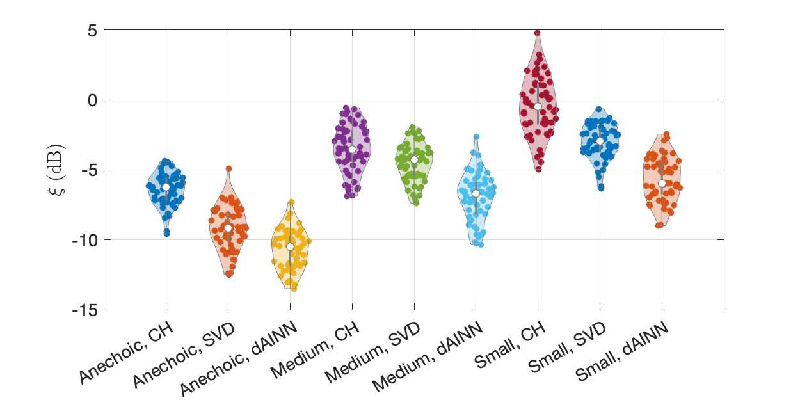}}
\caption{(color online) Pressure reconstruction error for all loudspeakers at 3 kHz in three rooms.
Small, CH denote the pressure reconstruction error of the CH 
method in the small room. Other labels can be interpreted similarly.
} 
\label{fig:figure8}
\end{figure}

The SFR experiment was repeated over all the 60 loudspeakers.  We set each loudspeaker to the planar array's transfer function as the 
ground truth, and used the sound pressure measured by the exterior 
28 microphones to reconstruct the sound pressure at the interior 
36 microphones within the planar array. Fig.~\ref{fig:figure8} 
compares the sound pressure reconstruction errors of the CH, SVD, and dAINN methods in the three rooms 
for all loudspeakers at 3 kHz. 

Fig.~\ref{fig:figure8} shows that the reconstruction errors in the anechoic chamber 
are smaller than those in other two rooms, as expected, due to the simpler pattern of the sound field. 
The dAINN achieves the smallest average reconstruction errors of -12.5, -8, and 
-7.2 dB, in the anechoic chamber, the medium room, and the small 
room, respectively. 
Similar experiments were also conducted at 1 kHz and 2 kHz. 
Except that the reconstruction errors were larger for the medium room 
and the small room at 1 kHz the results were similar to Fig.~\ref{fig:figure8},
thus are not shown here for brevity. 

\subsection{Pressure gradient reconstruction: loudspeaker 7\label{sec:gradient_7}}

\begin{figure*}[t]
\centerline{\includegraphics[width=17cm]{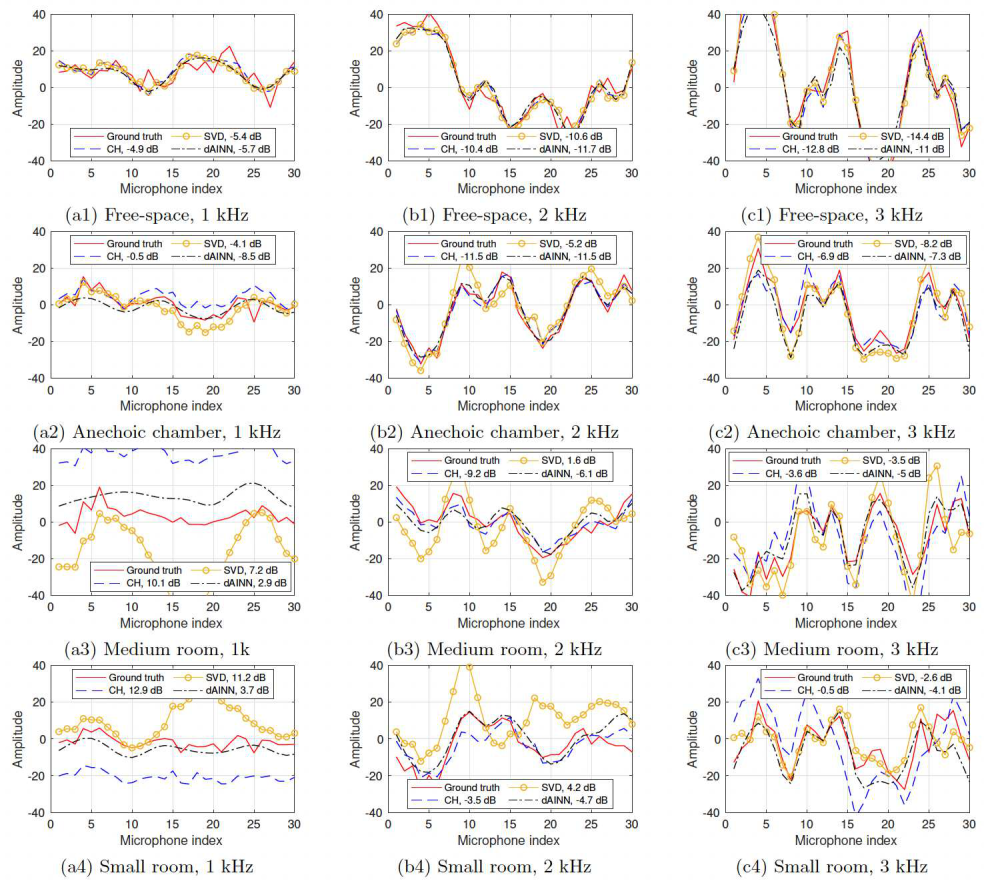}}
\caption{(color online) Pressure gradient reconstruction:
reconstruct the radial pressure gradient based the measurement of 30 microphones 
in the exterior circle of the dual-circular array - Loudspeaker 7.
The numbers in each sub-figures denote the 
reconstruction errors by the SH, SVD, and dAINN methods.
} 
\label{fig:figure9}
\end{figure*}
The sound pressures measured by the 30 microphones on the exterior 
circle of the dual-circular array (Fig.~\ref{fig:figure3} (c)) were used to reconstruct the radial pressure gradients between the two circles. 
The ground truth radial pressure gradients could not be directly measured by the set up shown in Figs.~\ref{fig:figure3} and \ref{fig:figure4}, 
thus were approximated as  
\begin{equation}
\label{eq:gradient_truth}
\frac{\partial{P}(r,\theta_m)}{\partial{r}} 
\approx \frac{
{P}(r_1,\theta_m)- 
{P}(r_2,\theta_m) 
}{r_1-r_2},  
\end{equation}
where $m=1, 2, ..., 30$.
The radial pressure gradients were reconstructed by the CH method, 
the SVD method and the dAINN method through Eq.~\eqref{eq:ch_rad_gradient},Eq.~\eqref{eq:svd_rad_gradient}, and Eq.~\eqref{eq:ainn_rad_gradient}, respectively.

Except using the measured sound pressures 
to reconstruct the pressure gradient, we conducted one more 
numerical experiment using simulated data. 
The transfer functions between loudspeaker 7 and the 
dual-circular array were simulated using the free-field 
Green's function~\cite{william1999} 
with added white Gaussian noise at a signal-to-noise (SNR) ratio of 20 dB 
to model disturbances.  
The reconstruction errors are shown in Fig.~\ref{fig:figure9}. 
As shown in Fig.~\ref{fig:figure9} (a1), (b1), and (c1), the 
reconstruction errors of three methods are relatively small, 
i.e., $<-10$ dB, when using the simulated 
sound pressures for reconstruction.

However, as shown in Fig.~\ref{fig:figure9} (a2) - (c4), when the measured sound pressures, 
which are expected to have more measurement noises,
are used, 
the reconstruction errors of all the three methods show signs of degradation. The data-dependent CH and SVD methods exhibit significant deviations from the ground truth in the medium room and the small room. In contrast, by exploiting the data-independent Helmholtz equation for regularization, the dAINN method 
is less susceptible to the inherent disturbances of the measurement processes. Thus, the dAINN method achieves the least pressure gradient reconstruction errors in most of the cases.

\begin{figure}[t]
\centering
\centerline{\includegraphics[width=9.6cm]{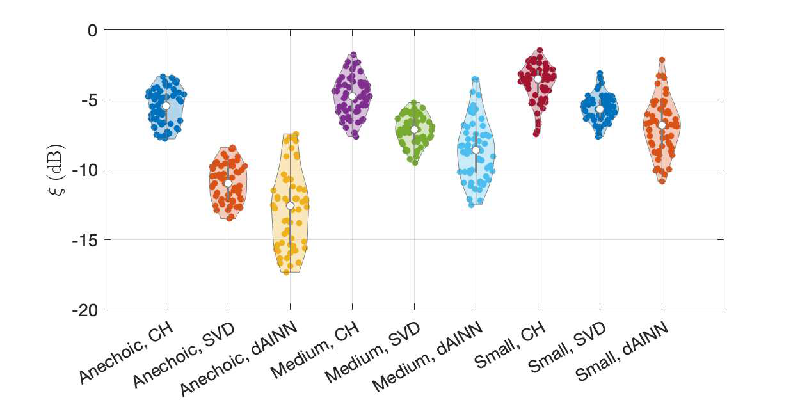}}
\caption{(color online) Pressure gradient reconstruction error for all loudspeakers at 3 kHz in three rooms.
Small, CH denote the pressure gradient reconstruction error of the CH 
method in the small room. Other labels can be interpreted similarly.
} 
\label{fig:figure10}
\end{figure}

\subsection{Pressure gradient reconstruction: all loudspeakers \label{sec:gradient_all}}
The pressure gradient reconstruction experiment was repeated over all 60 loudspeakers.
We set each loudspeaker to the dual-circular array's transfer function
as the sound pressure, approximated the radial pressure gradient 
according to Eq.~\eqref{eq:gradient_truth}, and reconstructed the 
pressure gradient in the same way as in Sec.~\ref{sec:gradient_7}.

Fig.~\ref{fig:figure10} shows reconstruction errors for all loudspeakers in the three rooms by the CH , SVD , and dAINN methods at 3 kHz. 
The experimental results for 1 and 2 kHz showed similar results 
as Fig.~\ref{fig:figure10}, thus are not shown for brevity.  
Comparing Fig.~\ref{fig:figure10} with Fig.~\ref{fig:figure8},
we can see that the pressure gradient is more challenging to reconstruct 
than the sound pressure. 
The dAINN achieves the smallest average reconstruction errors 
of -11.5, -7.0, and -5.5 dB, in the anechoic chamber, the medium room, 
and the small room, respectively. 
The average reconstruction errors of the SVD method are about 2 dB higher than those of the dAINN. 
The reconstruction errors of the CH method are the worst, which exceed 0 dB in the small room.

It is noted that, in this paper, the pressure gradient was approximated by the finite difference,   Eq.~\eqref{eq:gradient_truth}.
This may not be accurate in some circumstances and may contribute to the relatively high 
reconstruction errors shown in Figs.~\ref{fig:figure9} and \ref{fig:figure10}. 
The velocity sensor~\cite{Vector2003} may be used for measuring the pressure 
gradient and testing the performance of the pressure gradient reconstruction in the future.

\section{Conclusion\label{sec:conclusion}}
This paper proposed a compact AINN method for SFR.  
A neural network was designed to approximate the measured sound pressure and 
obey the Helmholtz equation, which regularized the network to generate 
physically valid 
output at and beyond the measurement positions. 
The performance of the AINN method was validated by sound pressure and pressure 
gradient reconstruction experiments, and outperformed both the CH and 
SVD methods.
An extension of this work is to exploit knowledge of the sound source(s) to further improve the performance of the AINN method; this will be our topic for future work. 

\bibliography{ref}
\end{document}